\documentstyle[preprint,aps,epsf]{revtex}
\begin{document}
\draft

\title{\bf Genuine Three-Body Bose-Einstein Correlations and Percolation of 
Strings}
 
\author{M. A. Braun}
\address{High-energy physics department\\
St. Petersburg University \\
198904 St. Petersburg, Russia\\
}
\author{F. del Moral and C. Pajares}
\address{Departamento de F\'{\i}sica de Part\'{\i}culas\\
Universidade de Santiago de Compostela\\
15706 Santiago de Compostela, Spain
}

\maketitle

\begin{abstract}
Recent data show 
a large difference of the genuine three-body Bose-Einstein correlations
in S-Pb collisions 
and in Pb-Pb central collisions being close to zero in the first case
and to one in the second one.
These results, unexpected from conventional approaches, are naturally 
explained by the percolation of colour strings produced in the collisions
and subsequent incoherent fragmentation of the formed clusters. 
\end{abstract}

\pacs{25.75Gz; 25.75.-q; 12.38.Mh}

The strength of the genuine three-particle Bose-Einstein correlations
can be measured by the weight factor $\omega$ introduced in \cite{ref1p}:
\begin{equation} \omega=\frac{\Big\{C_3(q)-1\Big\}-\Big\{C_2(q_{12})-1\Big\}
-\Big\{C_2(q_{23})-1\Big\}-\Big\{C_2(q_{31})-1\Big\}}
{2\sqrt{\Big\{C_2(q_{12})-1\Big\}\Big\{C_2(q_{23})-1\Big\}
\Big\{C_2(q_{31})-1\Big\}}}\label{eq1}\end{equation}
Here $C_2$ and $C_3$ are the two- and three-body correlation functions
\cite{ref1pp} and
\begin{equation} q_{ij}=q_i-q_j,\ \ q^2=q_{12}^2+q_{23}^2+q_{31}^2\end{equation}
The weight-factor $\omega$ has been experimentally studied in 
e$^+$e$^-$ collisions by L3 collaboration, with the result consistent with
$\omega=1$ \cite{ref1}. It has also been studied in heavy-ion collisions.
NA44 collaboration \cite{ref1pp,ref2} has obtained $\omega=0.20\pm 0.02\pm 0.19$ for 
SPb collisions, i.e. compatible with no genuine three-body correlations.
On the other hand, the same
experiment with the same methodology has found $\omega=0.85\pm0.02\pm0.21$
for central Pb-Pb collisions. This value is compatible with
$\omega=0.606\pm0.005\pm0.178$ earlier reported by WA98 collaboration \cite{ref3}.

It has been recently suggested \cite{ref4} that an explanation of these data 
seems to be in line with the behaviour of the chaoticity parameter
$\lambda$ which measures the strength of the two-body Bose-Einstein
correlations
\begin{equation}\lambda=C_2(q=0)-1\label{eq3}\end{equation}
The intercept $\lambda$ is the most poorly determined parameter due to
Coulomb correction, dependence on the shape of correlation functions fits
and other uncertainties but 
experimental data in heavy-ion collisions show that for moderate atomic 
numbers of colliding nuclei $\lambda$ decreases with atomic number, as 
expected from the corresponding increase of the number of independent 
incoherent sources \cite{ref5,ref6}. Indeed going from O-C to O-Cu, O-Ag and O-Au,
$\lambda$
falls from 0.79 to 0.32 \cite{ref7}. However for heavier nuclei $\lambda$ no longer
decreases and eventually starts to increase. For S-Pb and Pb-Pb collisions 
NA44 obtains $\lambda=0.56$ and 0.59 respectively \cite{ref8}. Similar values 
have been found at RHIC for Au-Au collisions at $\sqrt{s}=130$ GeV \cite{ref9}.

This behaviour can be understood assuming that in a collision colour 
strings are formed stretched between the projectile and target, which then
break due to formation of quark-antiquark pairs. 
Each colour string is assumed to have a finite transverse dimension of area
$S_1=\pi r_0^2$ ($r_0\simeq 0.2$ fm). As the energy and/or atomic number of 
the projectile and target increase the number and density of strings grows,
so that they start to overlap, forming clusters, which act as new 
effective sources of particle production.
Both the strings and their clusters can be assumed  to be totally 
chaotic sources with $\lambda=1$ \cite{ref10}. 
Assuming that for 
particles coming from different strings there are no Bose-Einstein 
correlations \cite{ref11} one then obtains \cite{ref12}
\begin{equation}\lambda=\frac{n_S}{n_T}\label{eq4}\end{equation}
where $n_S$ and $n_T$ are the average numbers of particle pairs produced 
in a given rapidity and transverse momentum range from the same cluster
and from all the clusters respectively.  

It is clear from (\ref{eq4}) that $\lambda$ decreases with the number of incoherent
sources (clusters). For very large
energies and/or atomic numbers the clusterization process will diminish the
number of independent sources (asymptotically to unity). As a consequence
the chaoticity parameter $\lambda$ will grow.

This approach may be realized in the 
framework of different scenarios depending on the assumed form of the
interaction between strings at close distances \cite{ref13,ref14}.
If the area of a cluster is formed by the geometrical sum of overlapping 
strings then a phase transition is observed when the string density $\eta$
reaches a critical value $\eta_c$ \cite{ref15}. This percolation phase transition
corresponds to the appearance of at least one cluster which spans the 
whole  interaction transverse area. The value of $\eta_c$ lies in the 
interval 1.17-1.5 depending on the form of the profile functions of the
colliding nuclei. Staying within this percolation scenario, the dynamics 
of the string interaction still admits different possibilities. The 
observed behaviour of $\lambda$ favours considering each cluster as a 
single string with a higher colour given by the vectorial sum of colours
of the overlappig strings times a factor which takes into account the 
degree of overlapping. As a result, the number of particles $\mu_n$ 
produced by a cluster of area $S_n$ formed by $n$ strings is given by
\begin{equation}\mu_n=\sqrt{\frac{nS_n}{S_1}}\mu_1\label{eq5}\end{equation}
where $\mu_1$ is the number of particles produced by a simple string.
In the case of total overlapping $S_n=S_1$ and $\mu_n=\sqrt{n}\mu_1$.
In the opposite case when strings just touch each other
$S_n=nS_1$ and one gets $\mu_n=n\mu_1$ as expected.

To calculate $\omega$ we have to know
\begin{equation}\lambda_3=C_3(q=0)-1.\end{equation}
Under the above mentioned assumptions, similarly to (\ref{eq3}), we have
\begin{equation}\lambda_3=5\frac{n'_S}{n'_T}\end{equation}
where now $n'_S$ and $n'_T$ are the average numbers of particle  
triplets produced 
in a given rapidity and transverse momentum range from the same cluster
and from all the clusters respectively.

A completely chaotic cluster of $n$ strings will produce
$(1/2)\mu_n^2$ pairs of particles and $(1/6)\mu_n^3$ triplets of 
particles with $\mu_n$ given by (\ref{eq5}). Summing over all formed clusters 
$i=1,2,...M$ one obtains $n_S$ and $n'_S$. The total number of particles
produced from all clusters is obviously
\begin{equation}\mu=\sum_{i=1}^M\mu_{n_i}\end{equation}
The total numbers of pairs and triplets are $(1/2)\mu^2$ and
$(1/6)\mu^3$ respectively. Thus we find
\begin{equation}\lambda=\frac{<\sum_{i=1}^Mn_iS_{n_i}/S_1>}
{<\Big(\sum_{i=1}^M\sqrt{n_iS_{n_i}/S_1}\Big)^2>},\ \ 
\lambda_3=5\frac{<\sum_{i=1}^M\Big(n_iS_{n_i}/S_1\Big)^{3/2}>}
{<\Big(\sum_{i=1}^M\sqrt{n_iS_{n_i}/S_1}\Big)^3>}\label{eq9}\end{equation}

\begin{figure}
  \centering\leavevmode
  \epsfxsize=5in\epsffile{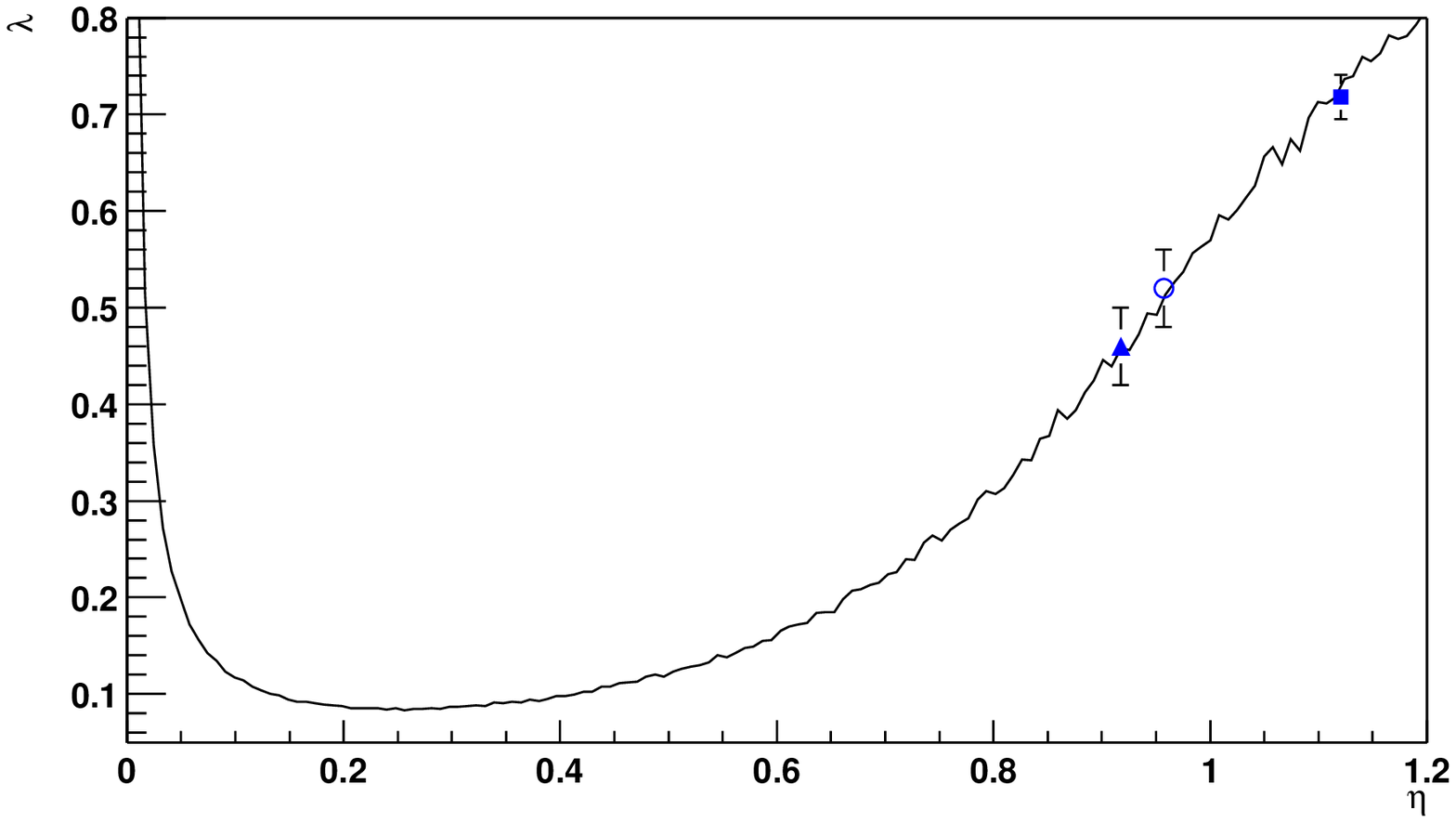}
\caption{The chaoticity parameter $\lambda$ as a function of $\eta$ from
Eq. (\ref{eq9}). The experimental points are for semi-central S-Pb
collisions [3] (filled triangle), 18\% central Pb-Pb
collisions [4] (nonfilled box) and 10\% central Pb-Pb
collisions [20] (filled box) at SPS.} \label{figure1}
\end{figure}

\begin{figure}
  \centering\leavevmode
  \epsfxsize=5in\epsffile{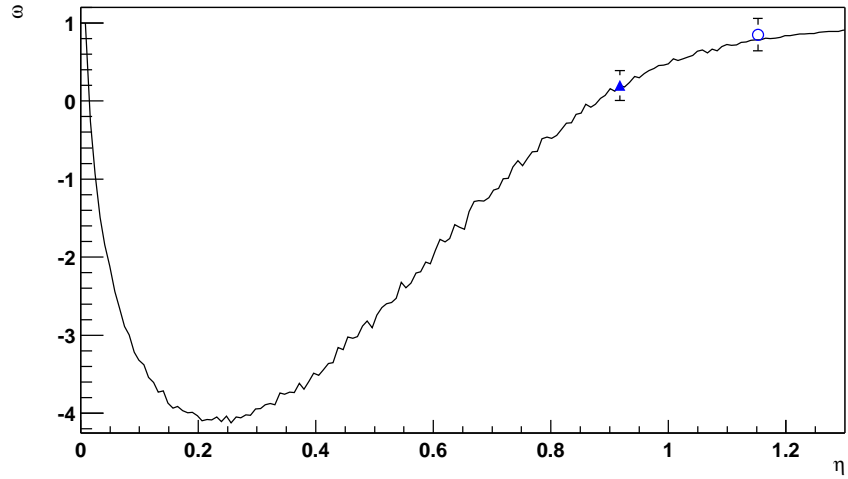}
\caption{The weight factor $\omega$ as a function of $\eta$ from Eqs. (\ref{eq1})
an (\ref{eq9}). The experimental points are for semi-central S-Pb
collisions [3] (filled triangle) and 9\% central Pb-Pb
collisions [4] (nonfilled circle) at SPS.}
\label{figure2}
\end{figure}

To calculate (\ref{eq9}) a Monte-Carlo simulation was performed. 
We generated $N$ discs of radius $r_0$ inside a circle of radius $R$
corresponding to the interaction area. For sufficiently large number of
discs (strings) and large $R$ both $\lambda$ and $\lambda_3$ result
depending only on the string density $\eta$ determined by
\begin{equation}\eta =\frac{Nr_0^2}{R^2}\end{equation}
Identifying all the formed clusters and determining
their areas we obtained $\lambda$ and $\omega$ 
as functions of $\eta$. Our results are presented in Figs. \ref{figure1} and \ref{figure2}. The
two experimental points of Fig. \ref{figure2} correspond to central S-Pb and Pb-Pb
collisions at SPS energies. The  number
of strings for these collisions is obtained from a
Monte-Carlo code \cite{ref17} based on the Quark-Gluon string model with
$r_0\simeq 0.2$ fm and $R$ corresponding to the required centrality. 

As to the chaoticity parameter $\lambda$, the obtained behaviour for it is
similar to our previous calculations in \cite{ref16}. Note that there a more 
elaborate approach was chosen in which we took into account 
energy-momentum conservation. The energy-momentum of each string was 
determined from that of the partons at its ends, which in its turn was 
given by the corresponding structure functions. Energy-momentum 
conservation limits the number of formed strings with energy
sufficient to produce particles. As a result, it shifts the minimum of the 
curve of Fig. \ref{figure1} to the right and makes its rise somewhat slowlier.

The dependence of $\omega$ on $\eta$ is found to be stronger than for 
$\lambda$, which was to be expected due to stronger dependence on the 
number of sources. This explains why the values of $\omega$ are
measured to be so different for S-Pb and Pb-Pb collisions at the same 
energy, and this is the reason of the good agreement obtained.

We expect that inclusion of energy-momentum conservation will only 
slightly modify the shape of the dependence of $\omega$ on $\eta$, as is 
the case of $\lambda$. Experimental information on $\omega$ at lower 
values of $\eta$ (e.g. from light nuclei or peripheral heavy ion 
collisions) would be most welcome to verify the predicted change of sign 
of $\omega$.

{\bf Acknowledgements}

This work has been done under contracts AEN99-0589-CO2 of CYCYT of Spain,
P61DTOOPX120613PN of Xunta de Galicia and NATO grant
PST.CLG.976799. M.A.B thanks the Secretaria de Estado de Educacion y 
Universidades of Spain for  the financial support.

\end{document}